\documentclass[conference]{IEEEtran}
\IEEEoverridecommandlockouts
\usepackage{cite}
\usepackage{amsmath,amssymb,amsfonts}
\usepackage{algorithmic}
\usepackage{graphicx}
\usepackage{textcomp}
\usepackage{xcolor}
\graphicspath{ {./images/} }
\usepackage{adjustbox}

\makeatletter
\newcommand{\linebreakand}{%
  \end{@IEEEauthorhalign}
  \hfill\mbox{}\par
  \mbox{}\hfill\begin{@IEEEauthorhalign}
}

\def\BibTeX{{\rm B\kern-.05em{\sc i\kern-.025em b}\kern-.08em
    T\kern-.1667em\lower.7ex\hbox{E}\kern-.125emX}}
\begin{document}

\title{Learning a faceted customer segmentation for discovering new business opportunities at Intel \\
}

\author{\IEEEauthorblockN{Itay Lieder}
\IEEEauthorblockA{\textit{Intel Advanced Analytics}}
\and
\IEEEauthorblockN{Meirav Segal}
\IEEEauthorblockA{\textit{Intel Advanced Analytics} }
\and
\IEEEauthorblockN{Eran Avidan}
\IEEEauthorblockA{\textit{Intel Advanced Analytics}}

\and
\IEEEauthorblockN{Asaf Cohen}
\IEEEauthorblockA{\textit{Intel Advanced Analytics}}
\linebreakand

\IEEEauthorblockN{Tom Hope}
\IEEEauthorblockA{\textit{Intel Advanced Analytics}}

}

\maketitle

\begin{abstract}

For sales and marketing organizations within large enterprises, identifying and understanding new markets, customers and partners is a key challenge. Intel’s Sales and Marketing Group (SMG) faces similar challenges while growing in new markets and domains and evolving its existing business. In today's complex technological and commercial landscape, there is need for intelligent automation supporting a fine-grained understanding of businesses in order to help SMG sift through millions of companies across many geographies and languages and identify relevant directions. 

We present a system developed in our company that mines millions of public business web pages, and extracts a faceted customer representation. We focus on two key customer aspects that are essential for finding relevant opportunities: industry segments (ranging from broad verticals such as healthcare, to more specific fields such as ``video analytics'') and functional roles (e.g., ``manufacturer'' or ``retail''). 

To address the challenge of labeled data collection, we enrich our data with external information gleaned from Wikipedia, and develop a semi-supervised multi-label, multi-lingual deep learning model that parses customer website texts and classifies them into their respective facets. Our system scans and indexes companies as part of a large-scale knowledge graph that currently holds tens of millions of connected entities with thousands being fetched, enriched and connected to the graph by the hour in real time, and also supports knowledge and insight discovery. In experiments conducted in our company, we are able to significantly boost the performance of sales personnel in the task of discovering new customers and commercial partnership opportunities.
\end{abstract}

\begin{IEEEkeywords}
AI for Enterprise, NLP, Web Mining
\end{IEEEkeywords}

\section{System Overview}

Our customer segmentation system is comprised of two major building blocks. The first component is tasked with large-scale data acquisition from the Web and other public sources, and consolidating it with internal corporate data in a Knowledge Graph (KG) we constructed. The second component is a suite of machine learning and natural language processing (NLP) models for segmenting potential customers.

\begin{figure}
  \includegraphics[width=\linewidth]{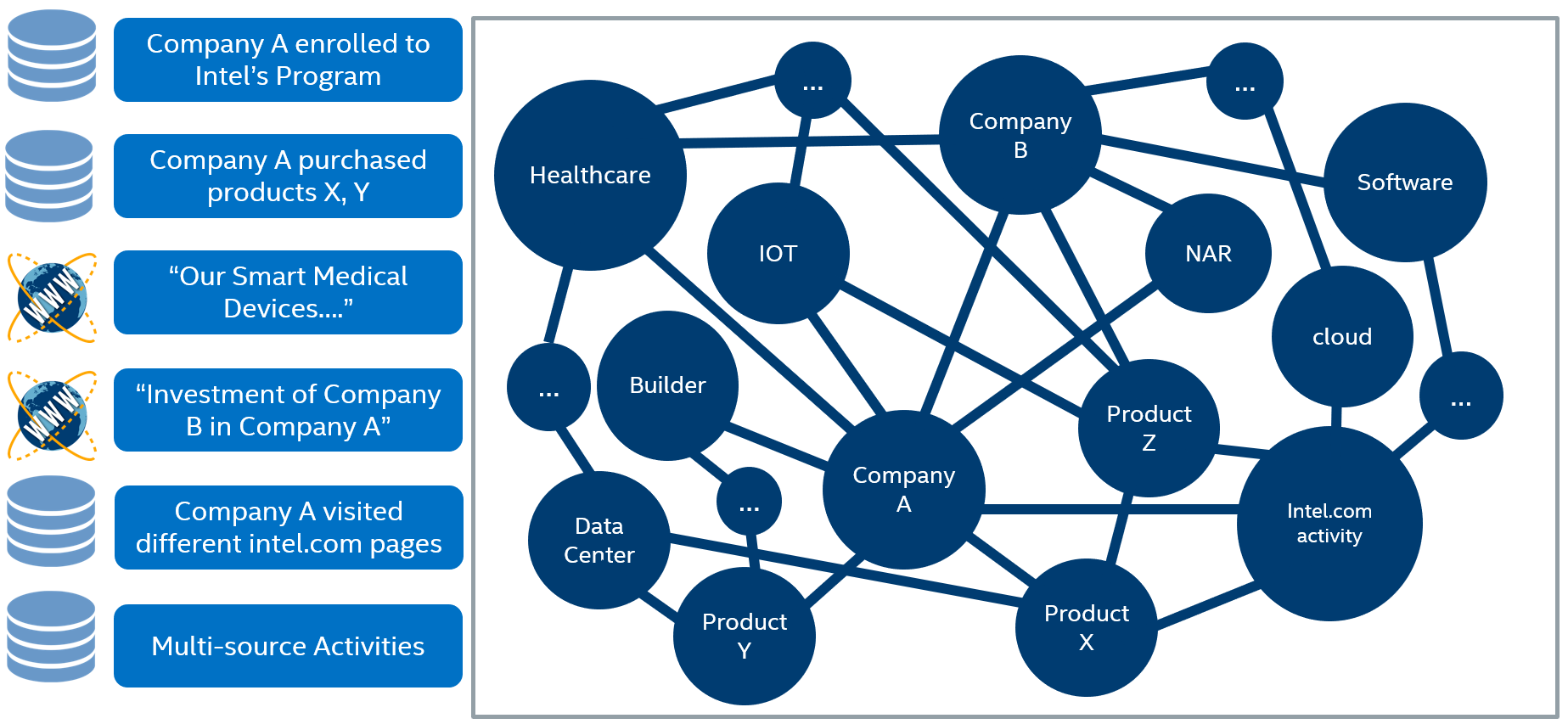}
    \caption{Illustration of our Knowledge Graph. Company information from public and internal sources.}
    \label{kg}
\end{figure}

\textbf{Large-scale knowledge graph and web crawling} Our solution requires constant streaming of textual data from millions of sites and updating of a multi-milllion node KG with gigabytes of data hourly (see Figure \ref{kg}).  We thus designed a parallelized, asynchronous streaming architecture using microservices to ensure robustness. We rely on Kafka for a message BUS, and use Kafka streams, TensorFlow serving and Neo4J. As existing customers often evolve and expand into new markets and nontraditional domains, our distributed and dynamic crawling process keeps web page information constantly refreshed, using cloud serverless architecture for scaling as the population grows. The data stream is transformed into graph formations using dedicated microservices and matched with the existing KG to assess whether new insertions or updates are warranted.

\textbf{Semi-supervised NLP model} Web pages are fed into a multi-label convolutional neural network (CNN) \cite{kim2014convolutional} text classification model, with a pre-trained multi-lingual BERT \cite{devlin2018bert} language model to help scale across languages and classes with scarce training data. 
We enrich our data by crawling tens of thousands of company sites appearing on Wikipedia with labeled industry information. We combine language and graph information to learn vector representations of industry concepts with a multi-view approach. Our model's page aggregation module focuses on key pages within each website and yields a unified prediction. For companies without labels we employ semi-supervised learning.

\section{Methods}
\subsection{Data} Training data was collected for companies either in business partnership with Intel (``internal'' data), or identified as relevant via information gathered from Wikipedia (``external'' data).

Our labels are comprised of two views, or facets, describing a company -- industries, and roles. In today's complex business ecosystem there are many industry segments, such as companies developing technological products and services in healthcare, education, transportation, communications, retail, and many more. Often, these segments can overlap or admit an hierarchical relationship. In many cases, companies work in several different fields and thus belong to multiple segments. In addition to the industries in which a company operates, another important facet is what role it plays -- a manufacturer, service provider, retailer, distributor, and more. 

A subset of internal companies was manually labeled by designated experts - Intel sales agents who closely monitor companies and may be in direct contact with them. The external companies are comprised of companies that appear in Wikipedia, and which we identified as relevant based on their industry, products, number of employees and more information which we extracted from their Wikipedia info-box. This method complemented our domain data with additional 77,323 companies, with a total number of 1,510,121 web-pages extracted and crawled from company websites. In addition to home pages, we select web-pages likely to be informative with respect to our task, as was implied by their URLs (e.g., \textit{"about\_us"}, \textit{"products"}, \textit{"technology")}.

\subsection{Model} \label{sec:model} Our primary model consists of five main components.

\textbf{Pre-trained multi-lingual language model} We leverage the multi-lingual BERT model, to capture contextualized representations of company texts, and scale across languages and sub-domains with more limited amounts of data. 

\textbf{Localized chunk representation} We decompose web-pages into chunks of text (based on HTML structure), each of limited length in order to fit the BERT architecture. We represent pages as a sequence of chunks, each a sequence of tokens. 

\textbf{Global pooling and prediction} We aggregate across the smaller chunks using a CNN with a multi-label objective. The convolutional aggregation is thus able to learn to identify informative parts of the page, and combine them in an adaptive, hierarchical manner to maximize segmentation accuracy. In addition to page-level pooling, we also pool across pages by computing the mean predicted probabilities across the respective individual pages.

We also experiment with fine-tuning BERT to our task, and with removing the CNN and directly training a linear classifier on top of BERT. Both these approaches were outperformed by our approach. We also use a CNN over pre-trained GloVe \cite{pennington2014glove} vectors without BERT, yielding competitive results.

\textbf{Semi-supervision} For external companies gleaned from Wikipedia, we had two types of labels: the first were labels of industries and products as they appear in Wikipedia, and the second were labels we inferred via an iterative semi-supervision process, where training data was comprised of the internal data (same subset used for training in all subsequent experiments to prevent over-fitting). 

\begin{figure}
\includegraphics[width=\linewidth]{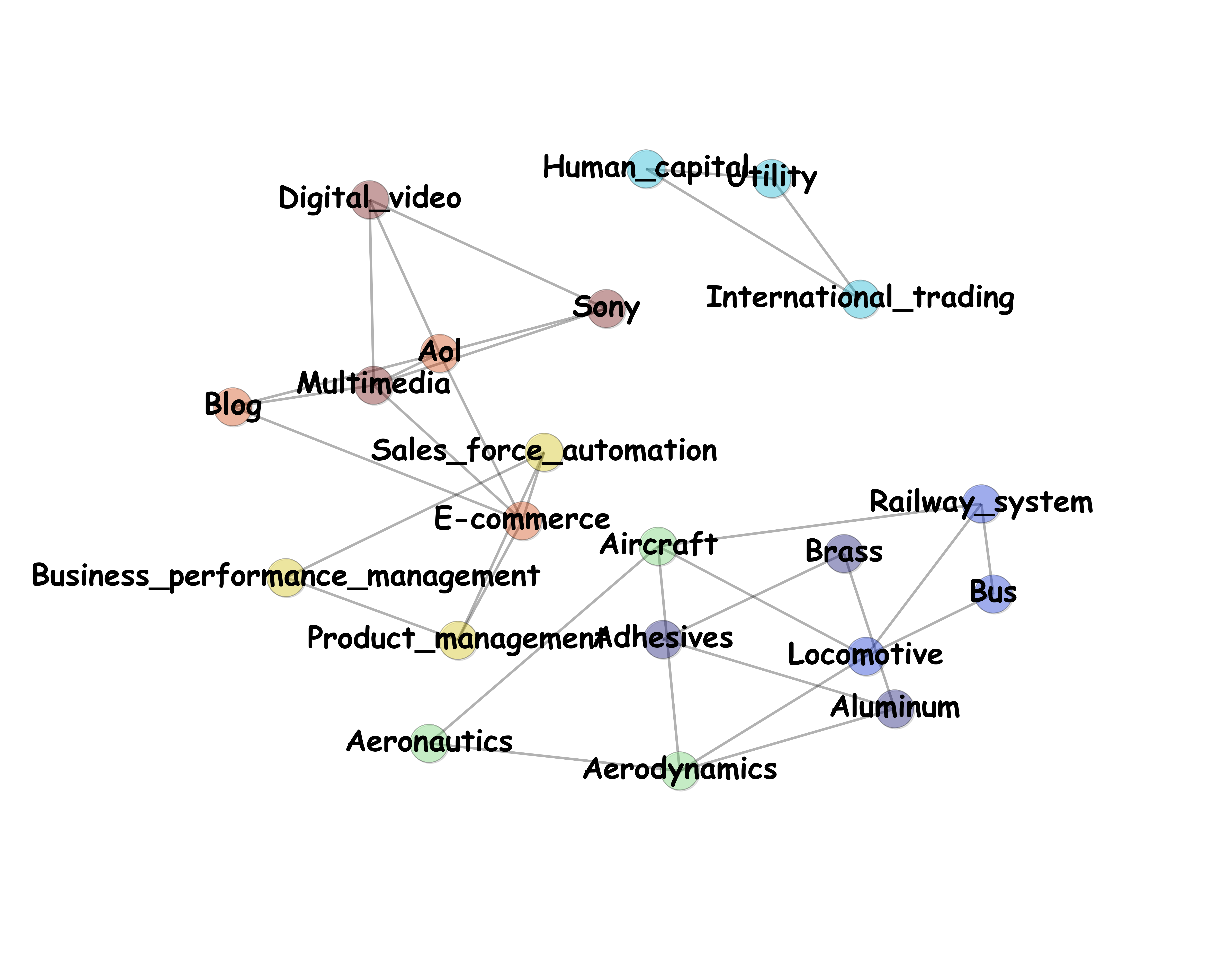}
\vspace*{-20mm}
    \caption{We combine language models and graph information to learn vector representations of industry concepts.}
    \label{fig:concept}
\end{figure}

\textbf{Label representation and aggregation} Due to a large label space with redundancies and hierarchical relationships (particularly for the industry facet), we seek to cluster labels into meaningful groups. We learn vector representations of industry segments and concepts, by fusing language model information and graph structures from Wikipedia. These representations are also used by SMG analysts to discover new links between companies and concepts (see Figure \ref{fig:concept}). Label clusters undergo human evaluation by SMG for final validation.

We compute relatedness among pairs of Wikipedia entities including companies, industries and products, with the following measures based on the Wikipedia graph (see \cite{ferragina2010tagme}): Milne-Witten, Jaccard, conditional probability, Barabasi-Albert, Pointwise Mutual Information, word2Vec text embedding, and the co-occurrence between industries and products. 
For each relatedness matrix, we applied Non-negative matrix factorization to deal with missing values. Finally, we regarded each of these relatedness measures as a different view and combined them into a single unified representation using Generalized Canonical Correlation Analysis (GCCA) \cite{benton2016learning}. See Figure \ref{fig:concept}, where edges are based on cosine similarity.

\section{Experiments} 

Training data included 70\% of the domains of the internal data, and evaluations were made on test data comprised of the complementary 30\% of the domains of the internal data. Also, each experiment was divided to two sub-experiments, with training and inference restricted to only one facet of labels - either industries or roles. This sub-division was motivated by empirically enhanced performance for each category of labels. 

To reduce noise, we kept only tokens with a minimum frequency of 10 appearances that also appear in the pre-trained GloVe vocabulary. Only pages with at least 20 clean tokens were included in the final training set.

For evaluation, we used the micro-averaged F1 and AUC scores, and additionally compared the F1 score at class-level. 

\textbf{Baselines} We compared our approach to several multi-label models in addition to the variants we discuss in the previous section (see Figure \ref{agg}).

\subsection{Experiment I} We explored the effect of adding external data (Wikipedia companies), labeled with semi-supervision, to our training set. We first trained our model only on internal data, then included external Wikipedia companies and compared performance.
 
\begin{figure}
  \includegraphics[width=\linewidth]{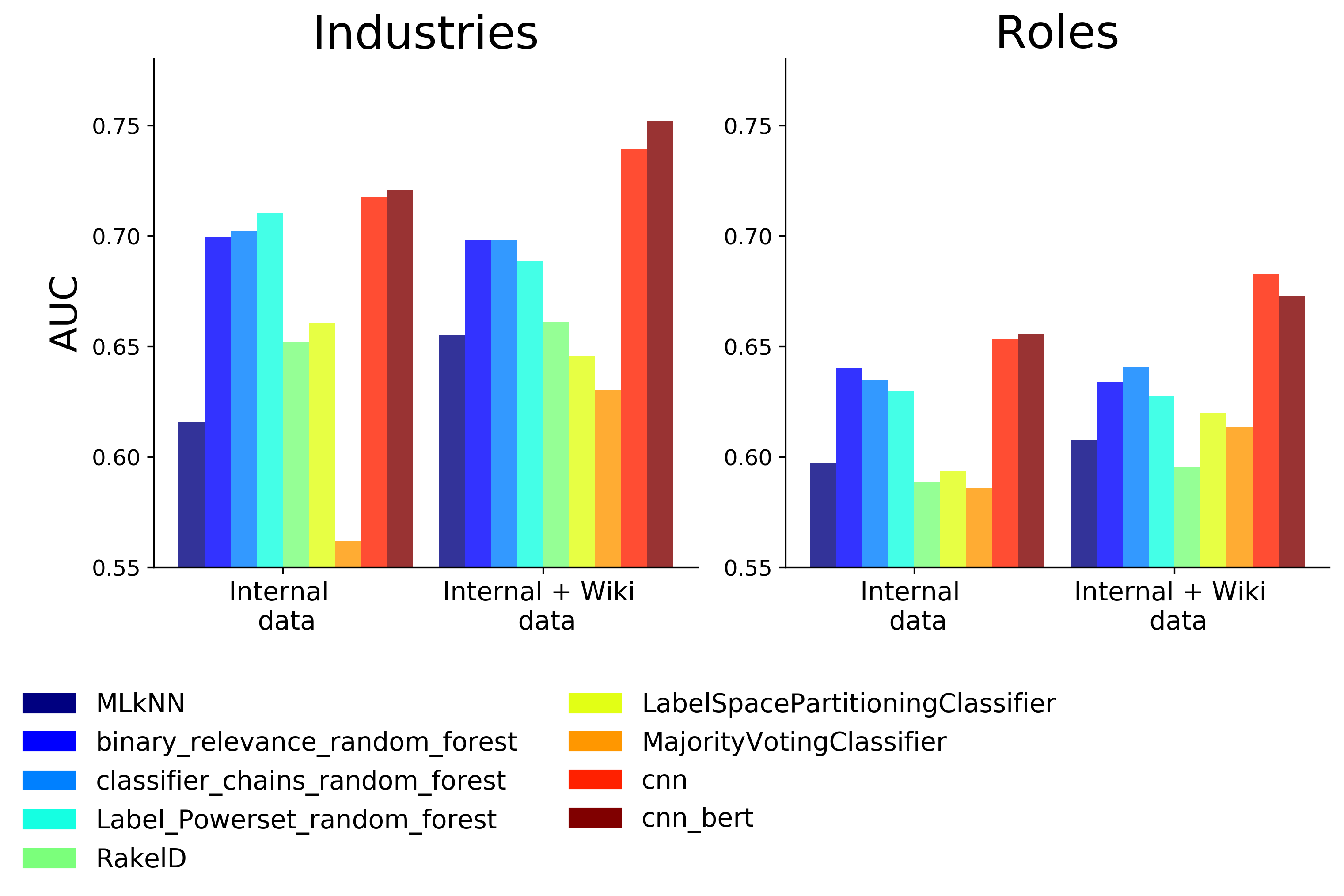}
    \caption{Micro-averaged AUC scores for each of the 9 tested models for industries (left) and roles (right). Left and right portions in each panel show performance without and with Wikipedia data. Best models were CNN with BERT (industries) and with GloVe (roles). Adding Wikipedia  improves performance in both cases.}\
\label{agg}
\end{figure}

\begin{figure}
  \includegraphics[width=\linewidth]{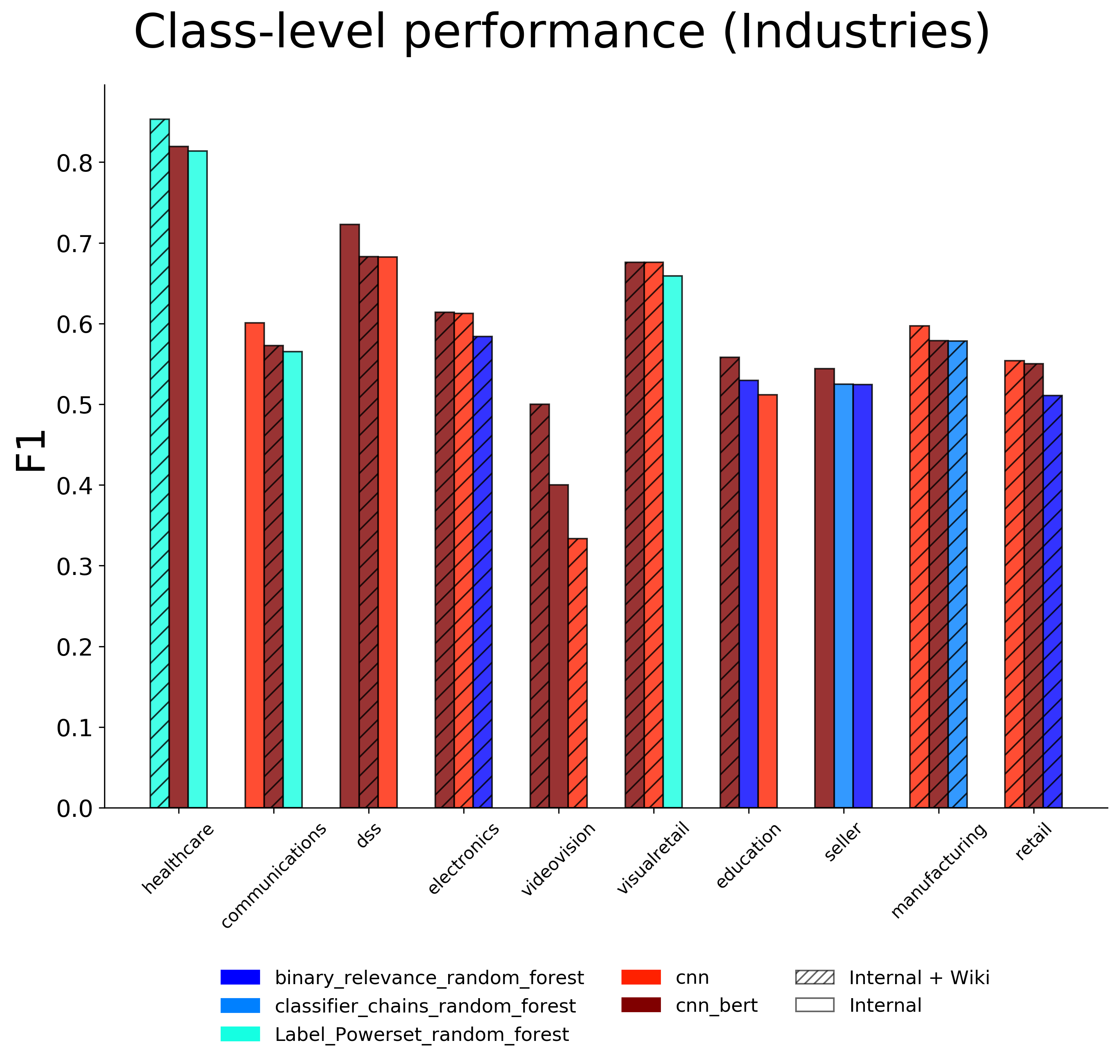}
    \caption{Class-level model comparison. For each of 7 representative industry segments and 3 representative roles, top 3 F1 scores (left-to-right) are shown. Hatch pattern denotes addition of Wikipedia data.}
\label{class}
\end{figure}

\textbf{Results} The two CNN models outperformed all other competitors by a large margin (Figure \ref{agg}). For industries, the CNN with BERT representations outperformed GloVe (F1=0.52 compared to 0.49). For roles, this trend was reversed (F1=0.44 compared to 0.39). In all cases, the best performing model was trained with the additional external data (F1=0.52 compared to 0.47 for industries, F1=0.44 compared to 0.39 for roles). 
On class-level (Figure \ref{class}), the two CNN models dominated for both industries (16/18 classes) and roles (6/8 classes). For industries, CNN with BERT was the superior model (outperforming on 11/18 classes), while using GloVe was best for predicting roles (4/8 classes). Adding external Wikipedia companies improved performance on 13/18 industry classes, and on 4/8 roles classes.

\textbf{Non-English languages} The multi-lingual BERT model had a larger impact on performance for languages other than English. In Spanish, for example, BERT boosted industry results (F1=0.44 compared to 0.32, AUC=0.74 compared to 0.65). For instance, BERT did better in correctly identifying the classes of \textit{consulting} (F1=0.42 compared to 0.11) and \textit{government} (F1=0.69 compared to 0.56).         

 \subsection{Experiment II} 
 
 For 2 specific classes -- \textit{"healthcare"}, \textit{"manufacturer"} -- we examined how performance in these classes is affected by replacing domains from the internal data with Wikipedia domains, keeping their original Wikipedia labeling. Our goal is to test the difference between internal Intel ``language'' and external concepts. For example, to examine the difference between the internally defined ``manufacturer'' concept (role) and the external Wikipedia-based concept. We hypothesized that for roles, which are based on more internal concepts, this effect should be more pronounced than for industries.

\textbf{Results} Indeed, only slight difference in performance was found between the two variants of the  \textit{"healthcare"} label, using data and labels from the internal versus external data (F1=0.76 compared to 0.73, 4\% increase). This was not true for the role label \textit{"manufacturer"}. Here performance was significantly better when training with the internal data (F1=0.5 compared to 0.31, 61\% increase). 

\section{Deployment results} Prior to our system, to find new leads SMG analysts used manual search, along with generic tools from external vendors that are not customized to internal language and needs (for example, not customized to SMG's \textit{manufacturer} concept). With our system, as part of a proof-of-concept experiment analysts discovered new leads for manufacturers (function) in specific industries with a large speed-up and a boost in accuracy in terms of leads found to be relevant and pursued.


\bibliographystyle{unsrt}
\bibliography{ref}

\end{document}